\documentclass{article}

\usepackage{arxiv}

\usepackage[utf8]{inputenc} 
\usepackage[T1]{fontenc}    
\usepackage{hyperref}       
\usepackage{url}            
\usepackage{booktabs}       
\usepackage{amsfonts}       
\usepackage{nicefrac}       
\usepackage{microtype}      
\usepackage{lipsum}		
\usepackage{graphicx}
\usepackage{natbib}
\usepackage{doi}
\usepackage{listings}
\usepackage[acronym]{glossaries}

\newacronym{sota}{SOTA}{state-of-the-art}
\newacronym{dpm}{DPM}{dynamic process management}
\newacronym{slurm}{SLURM}{Simple linux utility for resource management}
\newacronym{mqss}{MQSS}{Munich quantum software stack}
\newacronym{qdmi}{QDMI}{quantum device management interface} 
\newacronym{hcqw}{HCQW}{hybrid classical-quantum workflow} 
\newacronym{hpc}{HPC}{high performance computing}
\newacronym{wlm}{WLM}{workload manager}
\newacronym{qec}{QEC}{quantum error correction}
\newacronym{qem}{QEM}{quantum error mitigation}
\newacronym{nisq}{NISQ}{noisy, intermediate-scale quantum} 
\newacronym{mpi}{MPI}{message passing interface}
\newacronym{api}{API}{application programming interface}
\newacronym{rpc}{RPC}{remote procedure call}
\newacronym{hhl}{HHL}{Harrow-Hassidim-Lloyd}
\newacronym{qpu}{QPU}{quantum processing unit}
\newacronym{cpu}{CPU}{central processing unit}
\newacronym{gpu}{GPU}{graphics processing unit}
\newacronym{fpga}{FPGA}{field-programmable gate array}
\newacronym{ghz}{GHZ}{Greenberger-Horne-Zeilinger}
\newacronym{mpmd}{MPMD}{multiple program, multiple data}
\newacronym{qasm}{QASM}{quantum assembly language}
\newacronym{isc}{ISC}{International Supercomputing Conference}
\newacronym{mtbf}{MTBF}{mean time between failures}

\title{SLURM Heterogeneous Jobs for Hybrid Classical-Quantum Workflows}

\author{ \href{https://orcid.org/0000-0003-1597-0811}{\includegraphics[scale=0.06]{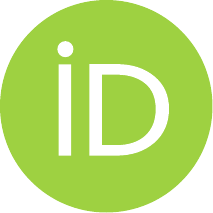}\hspace{1mm}Aniello Esposito}\thanks{Corresponding author.} \\
	EMEA Research Lab, HPE Labs\\
	Hewlett Packard Enterprise\\
	Basel, Switzerland \\
	\texttt{aniello.esposito@hpe.com} \\
	\And
	\href{https://orcid.org/0000-0001-7292-9984}{\includegraphics[scale=0.06]{orcid.pdf}\hspace{1mm}Utz-Uwe Haus} \\
	EMEA Research Lab, HPE Labs\\
	Hewlett Packard Enterprise\\
	Zurich, Switzerland \\
	\texttt{utz-uwe.haus@hpe.com} \\
}

\renewcommand{\headeright}{}
\renewcommand{\undertitle}{}

\hypersetup{
pdfauthor={Aniello Esposito},
pdfkeywords={},
}

\begin{document}
\maketitle

\begin{abstract}
A method for efficient scheduling of hybrid classical-quantum workflows is presented, based on standard tools available on common supercomputer systems. Moderate interventions by the user are required, such as splitting a monolithic workflow in to basic building blocks and ensuring the data flow. This bares the potential to significantly reduce idle time of the quantum resource as well as overall wall time of co-scheduled workflows. Relevant pseudo-code samples and scripts are provided to demonstrate the simplicity and working principles of the method. 
\end{abstract}


\section{Introduction}
Quantum computers offer the potential to efficiently tackle certain computationally hard problems with moderate input sizes, but their accessibility and ease of use remain limited. In contrast, supercomputers excel in data-intensive tasks and benefit from decades of development of mature and well-supported tool-chains, making them generally more accessible. Over the past decades, many promising quantum algorithms have been proposed~\cite{montanaro2016quantum}. Nevertheless, as discussed in \cite{mohseni2025buildquantumsupercomputerscaling} and \cite[Sec. 1.1]{davenport2023practical}, achieving parallelism in quantum algorithms poses significant challenges. Given the current limitations of early \gls{nisq} hardware, a more practical strategy would be to offload specific portions of classical workloads where quantum speedup is most effective. Numerous scientific applications~\cite{kiser2024contextualsubspaceauxiliaryfieldquantum,awsafqmc,esposito2024hybrid} could benefit from such hybrid execution models but standardized practices for integrating quantum and classical computation are still emerging~\cite{schulz2023towards,wille2024qdmi}. It is thus important for computational scientists to follow these developments but also understand what is already possible with the current ecosystem.
A hybrid system could be treated similarly to current multi-node HPC systems that integrate various accelerators like \gls{gpu} or \gls{fpga}. This compatibility implies that established tools like \gls{slurm}~\cite{slurm,yoo2003slurm} and \gls{mpi}~\cite{mpi40standard} can be adapted for hybrid environments. \gls{slurm}, for instance, already supports heterogeneous job scheduling and can be configured to implement \gls{hcqw} and \gls{mpi} provides \gls{dpm} which allows to disconnect a client from a server and thus releasing a quantum device as soon as possible. 

In this work we further develop the ideas presented in~\cite{esposito2023hybrid}. A hybrid job requiring repeated access to a quantum device is split into multiple sub-jobs which can be interleaved by \gls{slurm} to reduce the idle time of the quantum device. The only intervention by the user is to ensure information transfer between the jobs, e.g. with checkpoint and restart, while the communication can be hidden in an appropriate \gls{api}. The next section starts with a reference architecture and how a monolithic hybrid job can be split such that the basic blocks can release the quantum resource asap. The codes and scripts shown in this work consist of pseudo-code and are incomplete, but experiments with working code have been conducted on an HPE-Cray EX supercomputer featuring the necessary software environment as a proof of concept. 
\section{Methods}
\subsection{Architecture}
\begin{figure}[h!]
	\centering
	\includegraphics[width=\textwidth]{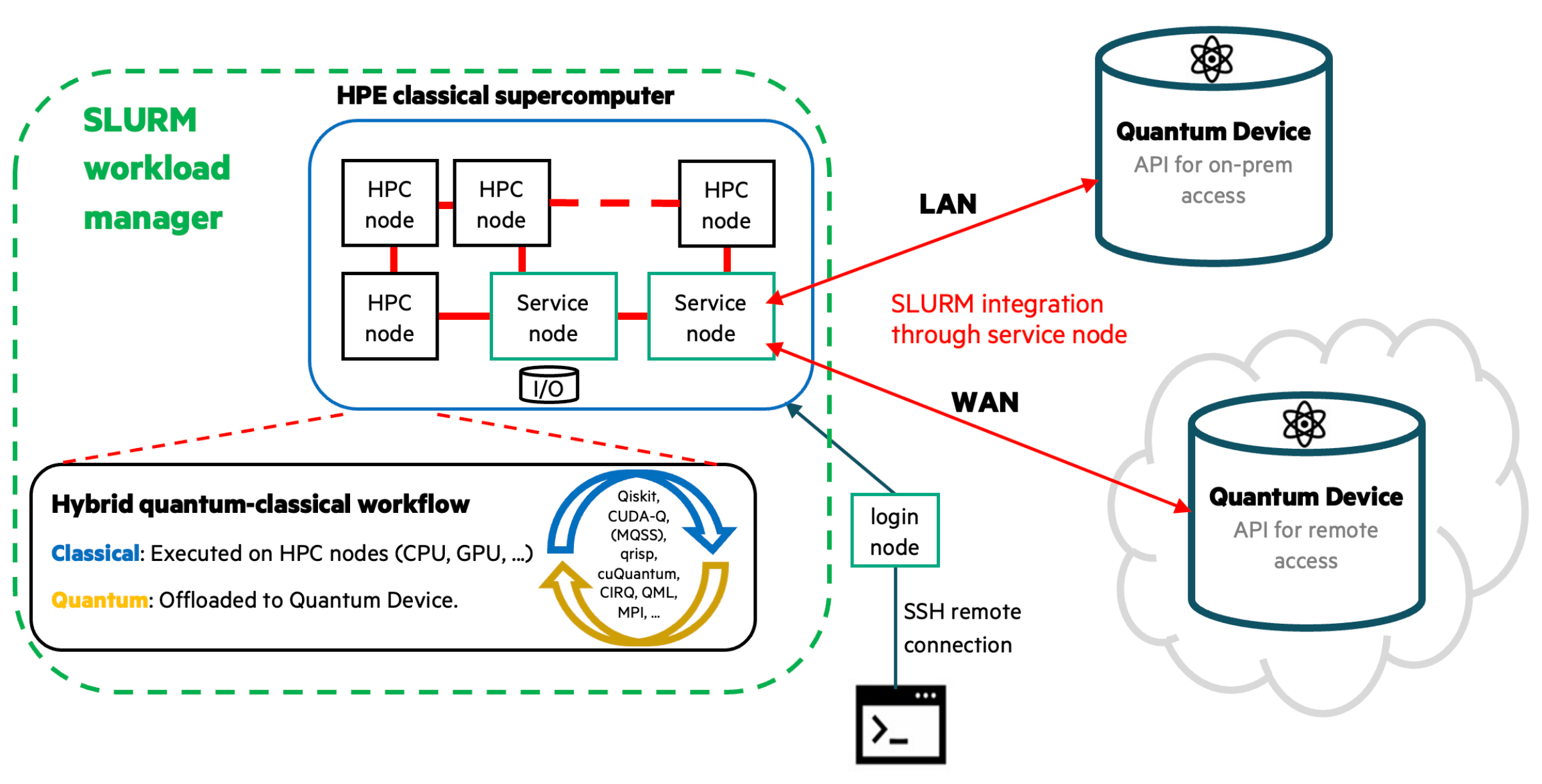}
	\caption{Basic architecture for the \gls{hcqw}s considered in this work. A dedicated service node provides exclusive access to a quantum device, either located in the same datacenter or via a cloud service. This service node is included in the resource pool of \gls{slurm}. Programs for the quantum device are generated with \gls{sota} tools.}
	\label{fig:architecture}
\end{figure}

A possible integration of a quantum device and a classical HPC cluster is shown in Fig.~\ref{fig:architecture}. A cluster based on classical hardware typically consists of several compute nodes connected through a high-speed fabric and a number of service nodes such as for storage. A quantum device can be connected exclusively to such a service node and communicate over different \gls{api}s, depending on its location, e.g. \texttt{http}. The \gls{slurm} workload manager encloses all relevant components, where the quantum device is exposed through the dedicated service node. Access to a service node as a separate resource is particularly advantageous in a multi-user environment. Popular \gls{sota} frameworks such as Qiskit~(\cite{wille2019ibm,javadi2024quantum,ckt}), CUDA-Q~(\cite{cuda-q}), Qrisp~(\cite{seidel2024qrisp}), Qsim~(\cite{isakov2021simulations,hancockcirq}), and Pennylane~(\cite{bergholm2018pennylane}) can be used to generate quantum programs, e.g. circuits, on the compute nodes and then sent to the service node for dispatching to the quantum device. The results returned from the device are then propagated by the service node to the relevant destinations, where communication is accomplished with \gls{mpi}. 

\subsection{Basic Workflow}
\begin{figure}[h!]
	\centering
	\includegraphics[width=0.75\textwidth]{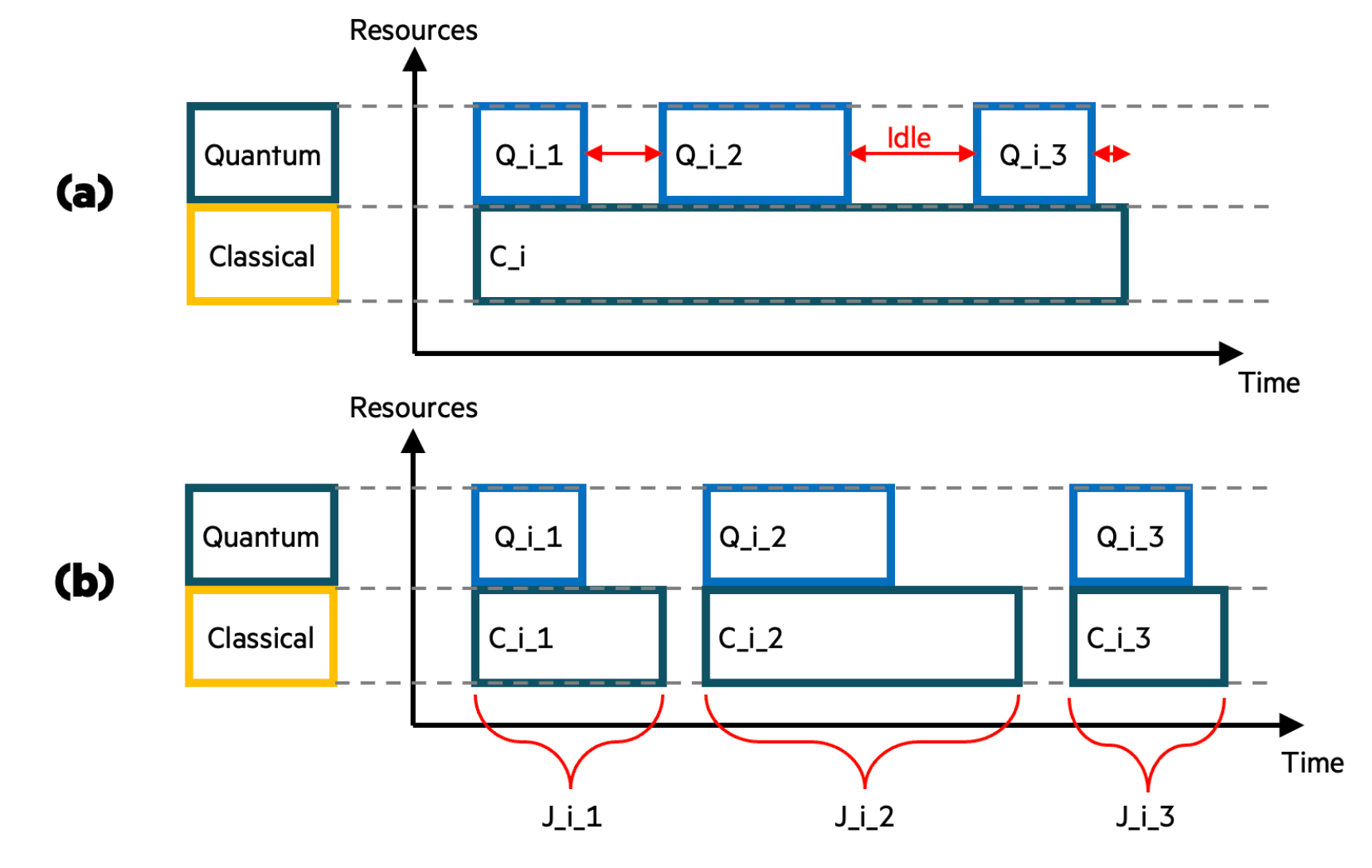}
	\caption{(a) \gls{hcqw} using a dedicated quantum device over the entire duration of the job. Red arrows indicate the potential idle time for the quantum device between the quantum blocks. Note that a simultaneous start of the classical and quantum part can always be achieved with an appropriate preparation. (b) Split of the workflow from (a) into smaller jobs containing only a single quantum block. The job identities are summarized, e.g. \texttt{C\_1\_1} and \texttt{Q\_1\_1} are combined into \texttt{J\_1\_1}.}
	\label{fig:job_split}
\end{figure}

The typical \gls{hcqw} considered in this work allocates a set of classical compute nodes for a contiguous amount of time and makes occasional use of a quantum device over a much shorter period as shown in Fig.~\ref{fig:job_split}.(a). The latter are referred to as quantum blocks \texttt{\{Q\_i\_1, Q\_i\_2, ...\}} and typically consist of more than just a single circuit or shot execution. Classical work is carried out throughout the duration of the classical block \texttt{C\_i} such as post-processing, communication to the quantum device, or the generation of new quantum programs. The allocation of heterogeneous resources like \gls{cpu}, \gls{gpu}, \gls{fpga}, or service nodes for access to quantum devices can be accomplished with \gls{slurm} heterogeneous jobs using an appropriate partition specifier \texttt{-p} as shown in listing~\ref{lst:slurm_single}. 
Communication between the components is accomplished with \gls{mpi} and a pseudo-implementation of an example program is shown in listing~\ref{lst:single}.
The item sent to the service node is typically an quantum program which is then dispatched to the quantum device for execution using the appropriate \gls{api}. The result is then sent back to the relevant destination. This program is repeated for all the quantum blocks in the same job as shown in listing~\ref{lst:full_job}.
Though, for a reasonable estimation of the total execution time, the quantum device should be ideally accessible immediately and therefore, a dedicated device needs to be allocated for roughly the same duration of the classical resources. Though, this can cause significant idle time of the quantum device especially when the blocks are small.  

\begin{minipage}[t]{\columnwidth}
\begin{lstlisting}[caption=\gls{slurm} Heterogeneous job script for \gls{hcqw} in Fig.~\ref{fig:job_split}.(a). The option \texttt{-p qpu} allocates a service node which is connected to a quantum device. A single \texttt{MPI\_COMM\_WORLD} is generated., label=lst:slurm_single, frame=tlrb]{Name}

    #!/bin/bash
    
    #SBATCH -N1
    #SBATCH -n 2
    #SBATCH -t 5
    #SBATCH -p cpu
    #SBATCH --exclusive
    #SBATCH hetjob
    #SBATCH -N1
    #SBATCH -n 1
    #SBATCH -p qpu
    #SBATCH --exclusive
    
    EXE='python single.py'
    
    srun  -u -l  $EXE : $EXE > out.$SLURM_JOBID.txt 2>&1 

\end{lstlisting}
\end{minipage}

\begin{minipage}[t]{\columnwidth}
\begin{lstlisting}[label=lst:single, caption=Single program executing both the classical and quantum component in a single \texttt{MPI\_COMM\_WORLD}. The last rank is located on the service noded for the quantum device. One could also use \texttt{revc} instead of \texttt{irecv} for overlapping work on the server side but with \texttt{irecv} one could potentially overlap large data transport from the client like parts of the statevector. ,frame=tlrb]{Name}

    # Send work item to quantum partition
    comm.Send([..., dest=size-1)
    
    # Post non-blocking receive for the result.
    req = icomm.Irecv(..., source=0)
    
    # do quantum work on the quantum ranks and overlapping classical work on 
    # the classical ranks.
    if rank == size-1:
        do_quantum_work() 
    else:
        do_classical_work() 
    
    # Wait for the quantum partition to return the result and use it.
    req.Wait()
    use_quantum_result(...)
    
    # Do more classical work and send more quantum work items to the
    # client if necessary.
    
    # Continue with remaining classical work 
    if rank != size-1:
        remaining_classical_work(...)

\end{lstlisting}
\end{minipage}

\noindent\begin{minipage}[t]{.48\columnwidth}
  \begin{lstlisting}[label=lst:full_job, caption=A possible code executed by the workflow shown in Fig.~\ref{fig:job_split}.(a). Listing~\ref{lst:single} shows a possible implementation of the single steps.,frame=tlrb]{Name}

# Perform N iterations of the same 
# pattern 
res = res_init
for _ in range(N):
    # Hybrid classical-quantum part with 
    # overlapping work. 
    h_res = hybrid_work(res)
    # Do remaining classical work while 
    # quantum device is idling
    res = remaining_classical_work(h_res)
    
  \end{lstlisting}
\end{minipage}\hfill
\begin{minipage}[t]{.48\columnwidth}
  \begin{lstlisting}[label=lst:split_job, caption=Sequence of possible code executed by the workflow Fig.~\ref{fig:job_split}.(b). The single steps could still be done like in listing~\ref{lst:single} using a single \texttt{MPI\_COMM\_WORLD} but now \gls{mpi} \gls{dpm} is used in a client server model as show in listings~\ref{lst:server} and \ref{lst:client} to allow timely release of the quantum resource. ,frame=tlrb]{Name}

# First program:
# Do hybrid and remaining classical work 
# followed by a checkpoint
h_res_1 = hybrid_work(res_init)
res_1 = remaining_classical_work(h_res_1)
ceckpoint(res_1)

# Second program:
# Read result from checkpoints and do
# next iteration
res_1 = read_checkpoint()
h_res_2 = hybrid_work(res_1)
res_2 = remaining_classical_work(h_res_2)
ceckpoint(res_2)

# Third program 
...
    
  \end{lstlisting}
\end{minipage}

\subsection{Splitting of Jobs and Optimized Scheduling}
\begin{figure}[h!]
	\centering
	\includegraphics[width=0.75\textwidth]{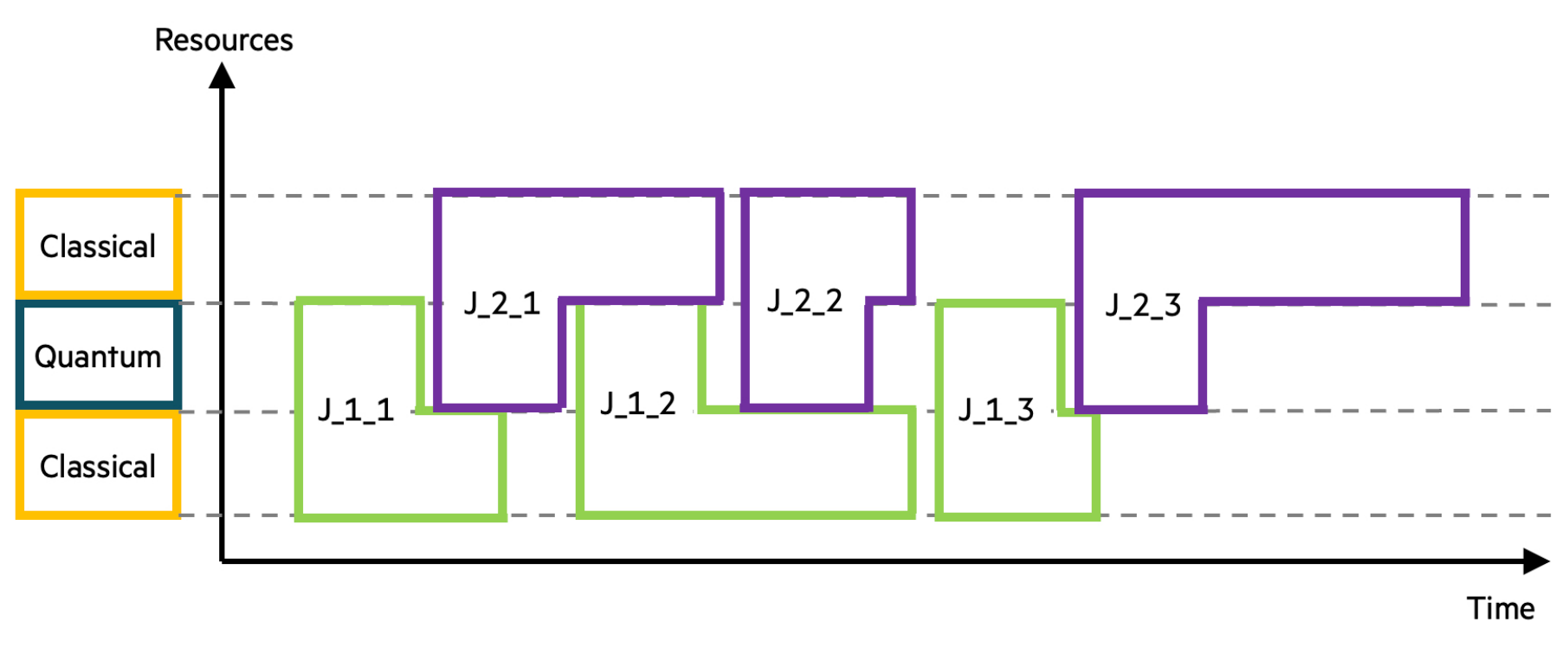}
	\caption{Optimized scheduling of two hybrid jobs which have been split in sub-jobs according to Fig.~\ref{fig:job_split}.}
	\label{fig:optimal_scheduling}
\end{figure}

In a multi-user environment several jobs of the form shown in Fig.~\ref{fig:job_split}.(a) are submitted to \gls{slurm}. If only a single quantum device is available, all jobs will be scheduled sequentially because of the required contiguous access to the quantum device. A possible reduction of the idle time begins with the split of the job in several sub-jobs containing only one quantum block as shown in Fig.~\ref{fig:job_split}.(b). Such a split requires the intervention of the code developer but can be straightforwardly achieved with simple methods such as checkpoint and restart, e.g. using \texttt{pickle} in \texttt{Python}. An example of such a code refactoring is shown in listing \ref{lst:split_job}.
The basic structure of the sub-jobs allows to release the quantum device right after the quantum block is done. For this purpose, a client-server model based on \gls{mpi} \gls{dpm} can be employed as shown in listings~\ref{lst:server} and \ref{lst:client}.
Every component has its own \texttt{MPI\_COMM\_WORLD} and the client job can be canceled by the server with a call to \texttt{scancel}.
These mechanisms can be embedded in an \gls{api} and made transparent to the user.
Releasing the quantum device as soon as possible, and thus saving precious resources, is a strong incentive for the user to invest time in these code modification.
The sub-jobs can then be submitted individually to \gls{slurm} specifying the dependency with the \texttt{-d} flag to \texttt{sbatch} to ensure the correct order. A possible batch script for \gls{dpm} in \gls{slurm} heterogeneous jobs is shown in listing~\ref{lst:slurm_split}
This gives an opportunity to \gls{slurm} to generate a scheduling as shown in Fig.~\ref{fig:optimal_scheduling} which reduces the overall idle time of the quantum device. This includes wasted quantum time allocated by the user as well as general idle time. 
Furthermore, the combined wall time of two jobs is reduced compared to a sequential scheduling.

\section{Related Work}
Research on \gls{hcqw}s is gaining significant momentum in the scientific computing community with a substantial number of research projects as well as publications and workshops at relevant \gls{hpc} conferences. The \gls{mqss}~\cite{schulz2023towards} based on the \gls{qdmi}~\cite{wille2024qdmi} is a promising example of such a long-term project. While this a sophisticated software stack for connecting end-users to the wide range of possible quantum devices, the present work aims to leverage the current \gls{sota} on supercomputer systems to bridge the gap until more advanced frameworks become available.

\section{Conclusion}
We showed that \gls{sota} tools available on common supercomputer systems such as \gls{slurm} and \gls{mpi} can be use to implement efficient \gls{hcqw}s. Splitting a monolithic hybrid job offers significant optimization potential in terms of reduction of quantum idle time and overall wall time of co-scheduled hybrid jobs. We have provided details of the implementation of the presented method as well as a scheme of a possible architecture. In a future work we want to collect data on a test system to statistically quantify the advantage of the proposed method. 

\section{Acknowledgment}
We want to thank the King Abdullah University of Science and Technology for the support through the advanced collaboration center for the Shaheen III HPE/Cray EX supercomputer. 

\noindent\begin{minipage}[t]{.48\columnwidth}
  \begin{lstlisting}[label=lst:server, caption=Server (classical component) with its own \texttt{MPI\_COMM\_WORLD},frame=tlrb]{Name}

# Create MPI Intercommunicator
MPI.Open_port(...)
MPI.Publish_name(...)
icomm = MPI.COMM_WORLD.Accept(...)

# Send work item to quantum device
icomm.Send([..., dest=0)

# Post non-blocking receive for the 
# result.
req = icomm.Irecv(..., source=0)

# Perform overlapping (classical) 
# work on the server side.
work(...)

# Wait for the quantum partition to 
# return the result and use it.
req.Wait()
use_quantum_result(...)

# Do more classical work and send 
# more quantum work items to the
# client if necessary.

# Shut down listener in client and 
# disconnect
shut_down_client_listener(...)
icomm.Disconnect(...)
MPI.Unpublish_name(...)
MPI.Close_port(...)

# Use scancel to terminate the client
# and therefore the quantum 
# part of the heterogeneous job
subprocess.run(["scancel", ...]) 

# Continue with remaining classical 
# work 
remaining_classical_work(...)
    
  \end{lstlisting}
\end{minipage}\hfill
\begin{minipage}[t]{.48\columnwidth}
  \begin{lstlisting}[label=lst:client, caption=Client (quantum component) with its own \texttt{MPI\_COMM\_WORLD},frame=tlrb]{Name}

# Create MPI Intercommunicator
MPI.Lookup_name(...)
icomm = MPI.COMM_WORLD.Connect(...)

# listener loop
while(...):
    # Receive work item from client
    icomm.Recv(..., source=0)
    
    # Execute work item on the quantum
    # device and send back.
    result = execute(...)
    icomm.Send(..., dest=0)

# Reach MPI.Finalize when listener 
# loop is shut down by server. 
    
  \end{lstlisting}
\end{minipage}

\begin{minipage}[t]{\columnwidth}
\begin{lstlisting}[caption=SLURM Heterogeneous job script for listing \ref{lst:full_job}. Every component will have its own \texttt{MPI\_COMM\_WORLD}. The option \texttt{-p qpu} allocates a service node which is connected to a quantum device. The \texttt{--network} options are relevant for \texttt{cray-mpich} on a HPE-Cray EX, label=lst:slurm_split, frame=tlrb]{Name}

    #!/bin/bash
    
    #SBATCH -N1
    #SBATCH -t 5
    #SBATCH -p cpu
    #SBATCH --network=single_node_vni,job_vni,def_tles=0
    #SBATCH --exclusive
    #SBATCH hetjob
    #SBATCH -N1
    #SBATCH -p qpu
    #SBATCH --network=single_node_vni,job_vni,def_tles=0
    #SBATCH --exclusive
    
    EXE1='python server.py'
    EXE2='python client.py'
    
    export MPICH_SINGLE_HOST_ENABLED=0
    export MPICH_DPM_DIR=${PWD}/dpm_dir
    
    srun  --het-group=0 -u -l -n 2 $EXE1 > server.$SLURM_JOBID.txt 2>&1 &
    sleep 2
    srun  --het-group=1 -u -l -n 1 $EXE2 > client.$SLURM_JOBID.txt 2>&1 
    wait

\end{lstlisting}
\end{minipage}

\bibliographystyle{unsrt}
\bibliography{references}  

\begin{thebibliography}{10}

\bibitem{montanaro2016quantum}
Ashley Montanaro.
\newblock Quantum algorithms: an overview.
\newblock {\em npj Quantum Information}, 2(1):1--8, 2016.

\bibitem{mohseni2025buildquantumsupercomputerscaling}
Masoud Mohseni, Artur Scherer, K.~Grace Johnson, Oded Wertheim, Matthew Otten, Navid~Anjum Aadit, Yuri Alexeev, Kirk~M. Bresniker, Kerem~Y. Camsari, Barbara Chapman, Soumitra Chatterjee, Gebremedhin~A. Dagnew, Aniello Esposito, Farah Fahim, Marco Fiorentino, Archit Gajjar, Abdullah Khalid, Xiangzhou Kong, Bohdan Kulchytskyy, Elica Kyoseva, Ruoyu Li, P.~Aaron Lott, Igor~L. Markov, Robert~F. McDermott, Giacomo Pedretti, Pooja Rao, Eleanor Rieffel, Allyson Silva, John Sorebo, Panagiotis Spentzouris, Ziv Steiner, Boyan Torosov, Davide Venturelli, Robert~J. Visser, Zak Webb, Xin Zhan, Yonatan Cohen, Pooya Ronagh, Alan Ho, Raymond~G. Beausoleil, and John~M. Martinis.
\newblock How to build a quantum supercomputer: Scaling from hundreds to millions of qubits, 2025.

\bibitem{davenport2023practical}
James~H. Davenport, Jessica~R. Jones, and Matthew Thomason.
\newblock A practical overview of quantum computing: Is exascale possible?, 2023.

\bibitem{kiser2024contextualsubspaceauxiliaryfieldquantum}
Matthew Kiser, Matthias Beuerle, and Fedor~Simkovic IV.
\newblock Contextual subspace auxiliary-field quantum monte carlo: Improved bias with reduced quantum resources, 2024.

\bibitem{awsafqmc}
Quantum monte carlo on quantum computers.
\newblock \url{https://github.com/amazon-braket/amazon-braket-examples/blob/feature/quantum-monte-carlo/examples/hybrid_quantum_algorithms/Quantum_Monte_Carlo_Chemistry/Quantum_Monte_Carlo_Chemistry.ipynb}.

\bibitem{esposito2024hybrid}
Aniello Esposito and Tamuz Danzig.
\newblock Hybrid classical-quantum simulation of maxcut using qaoa-in-qaoa.
\newblock In {\em 2024 IEEE International Parallel and Distributed Processing Symposium Workshops (IPDPSW)}, pages 1088--1094. IEEE, 2024.

\bibitem{schulz2023towards}
Martin Schulz, Laura Schulz, Martin Ruefenacht, and Robert Wille.
\newblock Towards the munich quantum software stack: Enabling efficient access and tool support for quantum computers.
\newblock In {\em 2023 IEEE International Conference on Quantum Computing and Engineering (QCE)}, volume~2, pages 399--400. IEEE, 2023.

\bibitem{wille2024qdmi}
Robert Wille, Ludwig Schmid, Yannick Stade, Jorge Echavarria, Martin Schulz, Laura Schulz, and Lukas Burgholzer.
\newblock Qdmi-quantum device management interface: Hardware-software interface for the munich quantum software stack.
\newblock In {\em 2024 IEEE International Conference on Quantum Computing and Engineering (QCE)}, volume~2, pages 573--574. IEEE, 2024.

\bibitem{slurm}
Slurm workload manager.
\newblock \url{https://slurm.schedmd.com/}.

\bibitem{yoo2003slurm}
Andy~B Yoo, Morris~A Jette, and Mark Grondona.
\newblock Slurm: Simple linux utility for resource management.
\newblock In {\em Workshop on job scheduling strategies for parallel processing}, pages 44--60. Springer, 2003.

\bibitem{mpi40standard}
{Message Passing Interface Forum}.
\newblock {\em {MPI}: A Message-Passing Interface Standard Version 4.0}, Jun 2021.

\bibitem{esposito2023hybrid}
Aniello Esposito, Jessica~R Jones, Sebastien Cabaniols, and David Brayford.
\newblock A hybrid classical-quantum hpc workload.
\newblock In {\em 2023 IEEE International Conference on Quantum Computing and Engineering (QCE)}, volume~2, pages 117--121. IEEE, 2023.

\bibitem{wille2019ibm}
Robert Wille, Rod Van~Meter, and Yehuda Naveh.
\newblock Ibm’s qiskit tool chain: Working with and developing for real quantum computers.
\newblock In {\em 2019 Design, Automation \& Test in Europe Conference \& Exhibition (DATE)}, pages 1234--1240. IEEE, 2019.

\bibitem{javadi2024quantum}
Ali Javadi-Abhari, Matthew Treinish, Kevin Krsulich, Christopher~J Wood, Jake Lishman, Julien Gacon, Simon Martiel, Paul~D Nation, Lev~S Bishop, Andrew~W Cross, et~al.
\newblock Quantum computing with qiskit.
\newblock {\em arXiv preprint arXiv:2405.08810}, 2024.

\bibitem{ckt}
Qiskit addon: circuit cutting.
\newblock \url{https://qiskit.github.io/qiskit-addon-cutting/}.

\bibitem{cuda-q}
Nvidia cuda-q.
\newblock \url{https://developer.nvidia.com/cuda-q#section-resources}.

\bibitem{seidel2024qrisp}
Raphael Seidel, Sebastian Bock, Ren{\'e} Zander, Matic Petri{\v{c}}, Niklas Steinmann, Nikolay Tcholtchev, and Manfred Hauswirth.
\newblock Qrisp: A framework for compilable high-level programming of gate-based quantum computers.
\newblock {\em arXiv preprint arXiv:2406.14792}, 2024.

\bibitem{isakov2021simulations}
Sergei~V Isakov, Dvir Kafri, Orion Martin, Catherine~Vollgraff Heidweiller, Wojciech Mruczkiewicz, Matthew~P Harrigan, Nicholas~C Rubin, Ross Thomson, Michael Broughton, Kevin Kissell, et~al.
\newblock Simulations of quantum circuits with approximate noise using qsim and cirq.
\newblock {\em arXiv preprint arXiv:2111.02396}, 2021.

\bibitem{hancockcirq}
Andrew Hancock, Austin Garcia, Jacob Shedenhelm, Jordan Cowen, and Calista Carey.
\newblock Cirq: A python framework for creating, editing, and invoking quantum circuits.

\bibitem{bergholm2018pennylane}
Ville Bergholm, Josh Izaac, Maria Schuld, Christian Gogolin, Shahnawaz Ahmed, Vishnu Ajith, M~Sohaib Alam, Guillermo Alonso-Linaje, B~AkashNarayanan, Ali Asadi, et~al.
\newblock Pennylane: Automatic differentiation of hybrid quantum-classical computations.
\newblock {\em arXiv preprint arXiv:1811.04968}, 2018.

\end{thebibliography}
%
%





\end{document}